\documentclass[epj]{svjour}
\usepackage{latexsym}
\usepackage{graphicx}
\usepackage{epstopdf}
\usepackage{amsmath}
\begin{document}
\title{Characterisation of a three-dimensional Brownian motor in optical lattices}
%\subtitle{Do you have a subtitle?\\ If so, write it here}
\author{P. Sj\"{o}lund\thanks{\email{peder.sjolund@physics.umu.se}} \and S. J. H. Petra \and C. M. Dion \and H. Hagman \and S. Jonsell\thanks{Current address: Department of Physics, University of Wales Swansea, Singleton Park, Swansea SA2 8PP, United Kingdom} \and A. Kastberg}
%\Offprints{peder.sjolund@physics.umu.se}
\institute{Department of Physics, Ume{\aa} University, SE-901 87 Ume\aa, Sweden}
\date{Received: date / Revised version: date}
\abstract{
We present here a detailed study of the behaviour of a three dimensional Brownian motor based on cold atoms in a double optical lattice [P. Sj\"{o}lund \textit{et al.}, Phys. Rev. Lett. \textbf{96}, 190602 (2006)]. This includes both experiments and numerical simulations of a Brownian particle. The potentials used are spatially and temporally symmetric, but combined spatiotemporal symmetry is broken by phase shifts and asymmetric transfer rates between potentials. The diffusion of atoms in the optical lattices is rectified and controlled both in direction and speed along three dimensions. We explore a large range of experimental parameters, where irradiances and detunings of the optical lattice lights are varied within the dissipative regime. Induced drift velocities in the order of one atomic recoil velocity have been achieved. 
} %end of abstract
\maketitle
\section{Introduction}
\label{Introduction}
Brownian motors (BMs) are small scale engines that convert random fluctuations into deterministic work. In order to realise such a device, symmetry has somehow to be broken (Curie's principle \cite{Curie}), and the system has to be out of thermodynamical equilibrium \cite{hanggi2005}. Much effort has been invested in studying the underlying mechanisms of BMs operating with the ratchet effect \cite{hanggi2005,Reimann,Astumian,Reimann2}. The ratchet mechanism consists in breaking the spatial and/or the temporal inversion symmetry of the system so that directed transport emerges, using fluctuations  as the relevant input. The paradigmatic device is Feynman's famous ratchet and pawl machine \cite{Feynman}, based on an idea of von Smoluchowski \cite{smo}. 

During the past years there has been an extensive body of work on BMs and ratchets, several realised in systems of periodic arrays of ultracold atoms, known as optical lattices (OLs) \cite{Mennerat,PederPRL,Renzoni,jones2007,carlo2006}. Optical lattices are periodic arrays of micro-traps created by the interference between two or more laser beams \cite{Jessen,grynberg01}. It has been shown that ultracold atoms stored in OLs can be controlled and manipulated with a very high degree of precision and flexibility and they are routinely used for example in studies of Bose-Einstein condensation and applicable for quantum state manipulation \cite{Bloch,Monroe}.

In this paper, we present an experimental and numerical study of fluctuations rectified into directed motion of Brownian particles in three dimensional OLs \cite{PederPRL}. This works by purely optical fields, where ultracold atoms are optically pumped between \emph{two} state-dependent and spatially overlapped OLs \cite{EllmannPRL,EllmannEPJD}, both spatially and temporally symmetric. These OLs are coupled by optical pumping, with strongly asymmetric transfer rates, \textit{via} the vacuum field reservoir, which, together with a shifted relative spatial phase between the OLs, causes atoms to be propelled in a controllable direction.

A general understanding of the dynamics of BMs is of fundamental interest. In biology, directed transport and molecular motors are both driven by stochastic motion, thus providing the energy input for the ability of living cells to generate motion and forces, \textit{e.g.}, for mobility, contraction of muscles or material transport, and are in themselves a fundamental mechanism of the origin of living cells. For example, biological motor proteins which move along linear filaments can be described by stochastic models coupled to chemical reactions \cite{Ebeling}. So-called ratchet models further explain the generation of directed motion on the microscopic level out of Brownian motion. The general concept of our BM, based on the idea of \cite{Laurent}, may be transferable into fields such as chemistry and biology as a tool for studies and control of, \textit{e.g.}, molecular motors, investigations of intra-cell motion and possible studies of Brownian motion in biological membranes \cite{Saffman}.
\section{Concept}
\label{sec:Concept}
The interaction between atoms and the interference pattern from the laser beams creates spatially periodic potentials (micro traps) seen by the atoms due to a second-order interaction between the atomic dipole and the light field \cite{Jessen,grynberg01}. Tuning the light frequency of an optical lattice close to an atomic resonance provides a dissipative channel resulting from spontaneous emission. The OL is then accompanied by an efficient cooling mechanism \cite{grynberg01,Chu,Cohen,Phillips}, providing friction in the system, and the presence of dissipation, albeit small, will result in a slow normal diffusion of the atoms in the lattice \cite{grynberg01,Laurent2,Hodapp}.

\begin{figure}
\centerline{\includegraphics[width=0.7\columnwidth]{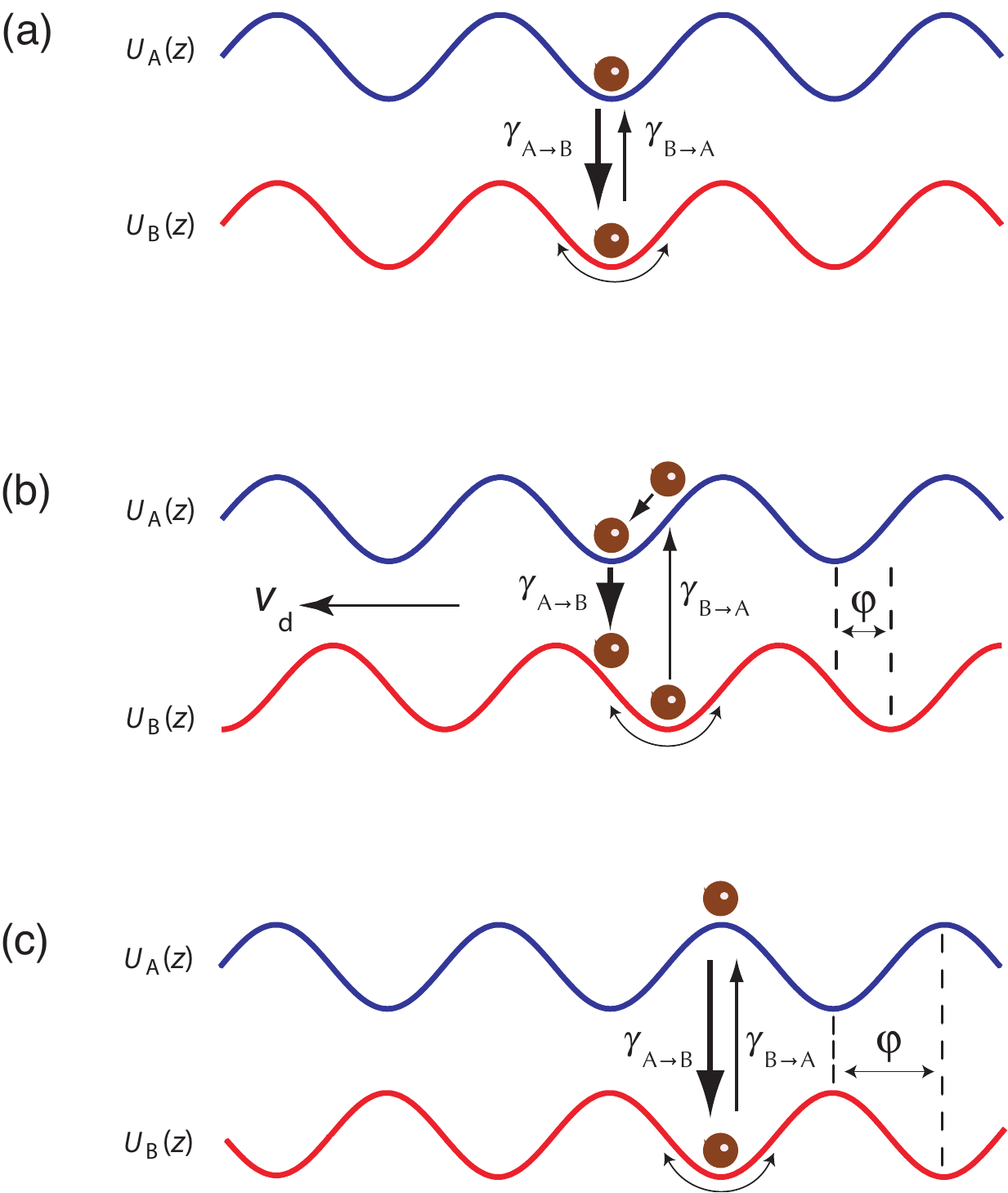}}
\caption{(Colour online). Rectification mechanism. Atoms move in two symmetric potentials $U_\mathrm{A}(z)$ and $U_\mathrm{B}(z)$ that are coupled with asymmetric optical pumping rates $\gamma_{\mathrm{A}\rightarrow \mathrm{B}}$ and $\gamma_{\mathrm{B} \rightarrow \mathrm{A}}$ ($\gamma_{\mathrm{A}\rightarrow \mathrm{B}}\gg \gamma_{\mathrm{B} \rightarrow \mathrm{A}}$). (a) The potentials are in phase. The transfer from the long-lived state B, to the transient state A, and back, does not lead to biased motion. (b) A relative phase shift $\varphi$ is introduced. Spatial diffusion is strongly facilitated in one direction, leading to a drift velocity $v_{\mathrm{d}_z}$. (c) The potentials are out of phase, $\varphi_z=\pi$, and no biased motion is introduced.}
  \label{fig:Concept}
\end{figure}

The basic idea and underlying principle of our BM is quite general \cite{Laurent}. Consider a Brownian particle stationary for a short time in one of two spatially and temporally symmetric sinusoidal potentials $U_\mathrm{A}(z)$ and $U_\mathrm{B}(z)$, see figure~\ref{fig:Concept}, and having a low kinetic energy compared to the modulation depths of the potentials.
This particle will undergo Hamiltonian motion around the potential minimum. The motion is interrupted by a dissipative process, resulting in a small overall Brownian motion, where the particle can be optically pumped between the two potentials. One of the components in the asymmetry that eventually gives rise to rectification is caused by unequal transfer rates between the potentials. Also, we assume the energy damping to be much smaller than the typical oscillation frequency in the potential wells, so that the dynamics falls into the low-damping regime, since no induced drift can be obtained for sinusoidal potentials in the overdamped regime \cite{Kanada}.

By letting the relative spatial phase between the two potentials be slightly shifted, combined with the pronounced difference in the transfer rates, the spatiotemporal symmetry of the system is broken and a directed motion is obtained. The concept can therefore be adapted to induce directed motion in any direction in three dimensions since the same physical process will function in any direction in a three-dimensional case. Because we use symmetric potentials, the direction of the induced drift can be reversed by a proper choice of the relative spatial phase shift. This is to be contrasted with the usual use of a ratchet potential (\textit{e.g.}, using a sawtooth potential), where the direction is fixed by the shape of the potential.  
\section{Experiment}
\label{sec:Experiment}
\subsection{Experimental setup}
The BM is realised, using cold caesium atoms in a double optical lattice (DOL) \cite{EllmannPRL,EllmannEPJD,PetraJOptA}. A magneto-optical trap (MOT) is loaded from a chirped-slowed atomic beam, produced in a thermal source which fills the MOT with approximately 10$^8$ atoms at a peak density of 10$^{11}$ cm$^{-3}$. Approximately 80\% of the initial number of atoms from an optical molasses of a few microkelvin is loaded into the DOL with a filling fraction of about 0.05 atom per site. After loading, the OL light is left on for a chosen time $\tau$. To avoid spurious spatial drifts of the atomic cloud caused by light pressure, great care has been taken to balance the laser beam power \cite{PetraJOptA}. Also, it is crucial that the relative spatial phase is kept constant.

\begin{figure}
\centerline{\includegraphics[width=0.9\columnwidth]{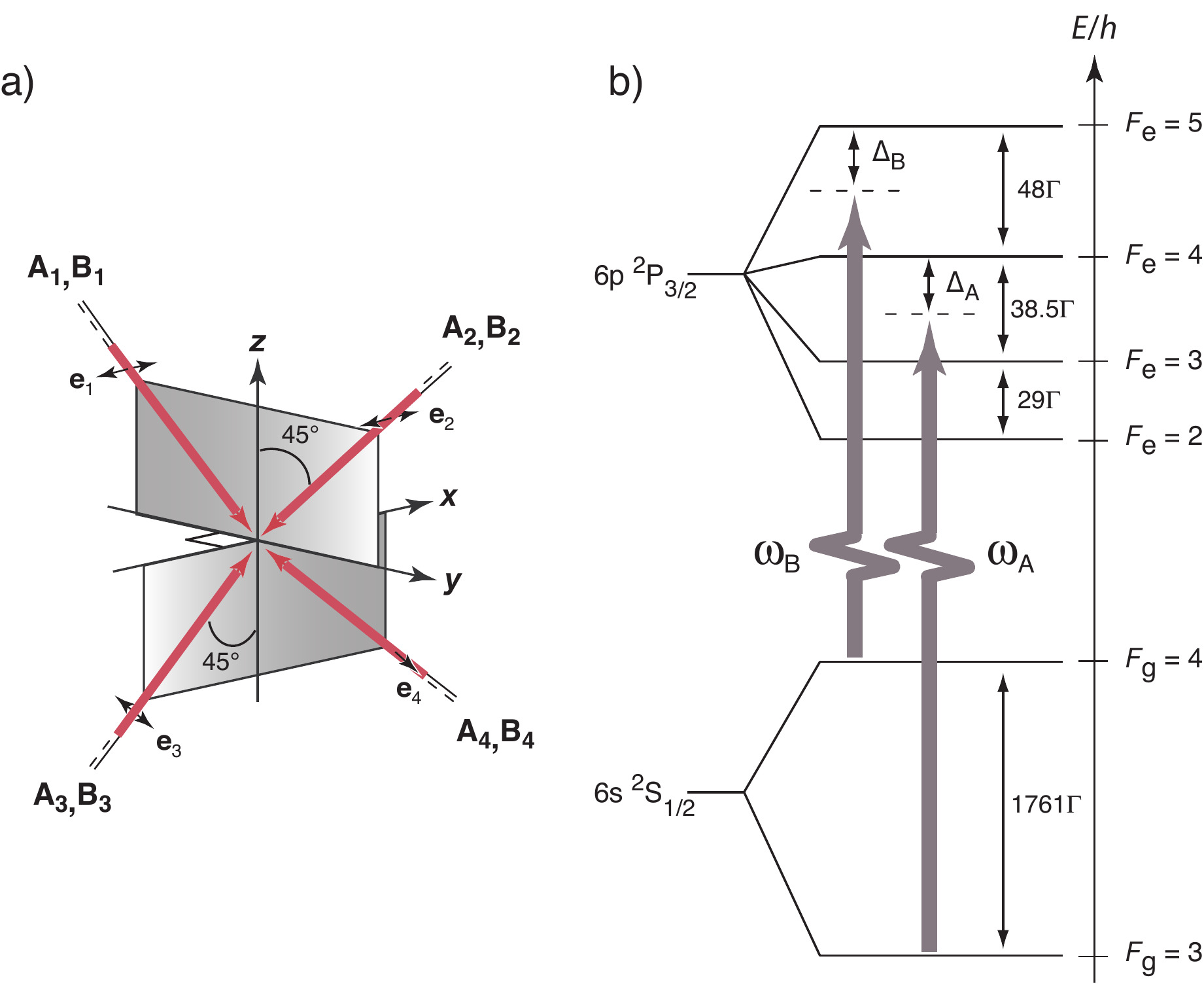}}
\caption{\textbf{a)} Double optical lattice beam geometry, realised by four branches, where each branch contains two spatially overlapped laser beams A and B. $\textbf{e}_1$ to $\textbf{e}_4$ are the polarisation vectors perpendicular to the plane of propagation. \textbf{b)} Schematic energy level diagram: Two ground states $F_\mathrm{g}=4$ and $F_\mathrm{g}=3$ are connected to the excited $F_\mathrm{e}=5$ and $F_\mathrm{e}=4$ states respectively by two laser fields B and A detuned by $\Delta_\mathrm{B}$ and $\Delta_\mathrm{A}$. $\Gamma/2\pi =5.22$ MHz is the natural linewidth of the transitions.}
\label{fig:levels}
\end{figure}

This three-dimensional DOL is composed of two spatially overlapped OLs with identical topography, but which can be controlled individually in terms of both well depths and relative spatial phase. Two different beams of frequencies $\omega_\mathrm{A}$ and $\omega_\mathrm{B}$ are split into four branches with equal power. This lattice geometry is a generalisation of the 1D lin$\perp$lin configuration to 3D \cite{grynberg01}, see figure \ref{fig:levels}a. We use frequencies that are near-resonant with the D2 line of caesium at 852 nm reaching the excited fine structure level 6p~$^2$P$_{3/2}$. Each optical lattice operates from a different hyperfine ground state of the 6s~$^2$S$_{1/2}$ level. One operates within the closed $F_\mathrm{g}=4\rightarrow F_\mathrm{e}=5$ transition, trapping atoms in the $F_\mathrm{g}=4$ ground state. The second lattice traps atoms in the $F_\mathrm{g}=3$ ground state, operating in the open $F_\mathrm{g}=3\rightarrow F_\mathrm{e}=4$ transition, see figure \ref{fig:levels}b. These resonances are much narrower than the difference in the laser frequencies, which allows us to address the two lattices independently. At the same time, the wavelengths are close enough, and the trapped atomic cloud is small enough (typically $1-2$ mm in diameter), to ensure that the periodicity is the same within the sample volume (\textit{e.g}., it takes about 3.3 cm for the lattices to phase out by $\pi$ in the horizontal $x$ and $y$-directions). 
Moreover, we can individually control the optical pumping rates between the lattices to some extent \cite{EllmannEPJD} by tuning the irradiances $I_\mathrm{A}$, $I_\mathrm{B}$ and the detunings $\Delta_\mathrm{A}$, $\Delta_\mathrm{B}$ of the lattice beams, which in turn changes the well depths, optical pumping rates, diffusion and friction, and consequently, also the behaviour of our BM.

By letting the two lattices be out of phase, \textit{e.g.}, in the vertical $z$-direction, while simultaneously having an asymmetric transfer rate between the lattices, the motion of atoms is channelled in that particular direction \cite{PederPRL}. This process can be controlled in all three dimensions, with regard to both speed and direction, by a proper choice of the relative spatial phases $\varphi_x$, $\varphi_y$, $\varphi_z$, irradiances $I_\mathrm{A}$ and $I_\mathrm{B}$ and the detunings $\Delta_\mathrm{A}$ and $\Delta_\mathrm{B}$ of the laser light.

The control of the relative spatial phase shifts between the lattices in three-dimensions is achieved by extending or shortening the lattice beam branches 1, 2, 3 and 4\footnote{This is done by manually adjusting the distance between two facing prisms, mounted on linear translation stages in each of the four branches.}, see figure \ref{fig:levels}a. A relative phase shift only in the horizontal $x$-direction corresponds to an extension of branch 3 and a shortening of branch 4 by the same amount, or vice versa. In $y$, an extension of branch 1 and shortening of branch 2 by the same amount, or the other way around, will cause a shift. For a phase shift only in the vertical $z$-direction, a simultaneous extension or shortening of both branches 1 and 2 or branches 3 and 4 by the same amount will work. Combining these, an arbitrary combination of $\varphi_x$, $\varphi_y$ and $\varphi_z$ can be set. The relative spatial phases between the lattices are experimentally determined by the kinetic temperature, measured by a ballistic time-of-flight (TOF) technique \cite{lett}, when changing the relative spatial phase, see figure \ref{fig:Temp}. A Gaussian fit to the TOF-signal also provides the peak arrival time of the distribution and the number of atoms by the area under the fitted curve.
\subsection{Calibration of the relative spatial phase by temperature measurement}
\begin{figure}
\centerline{\includegraphics[width=0.7\columnwidth]{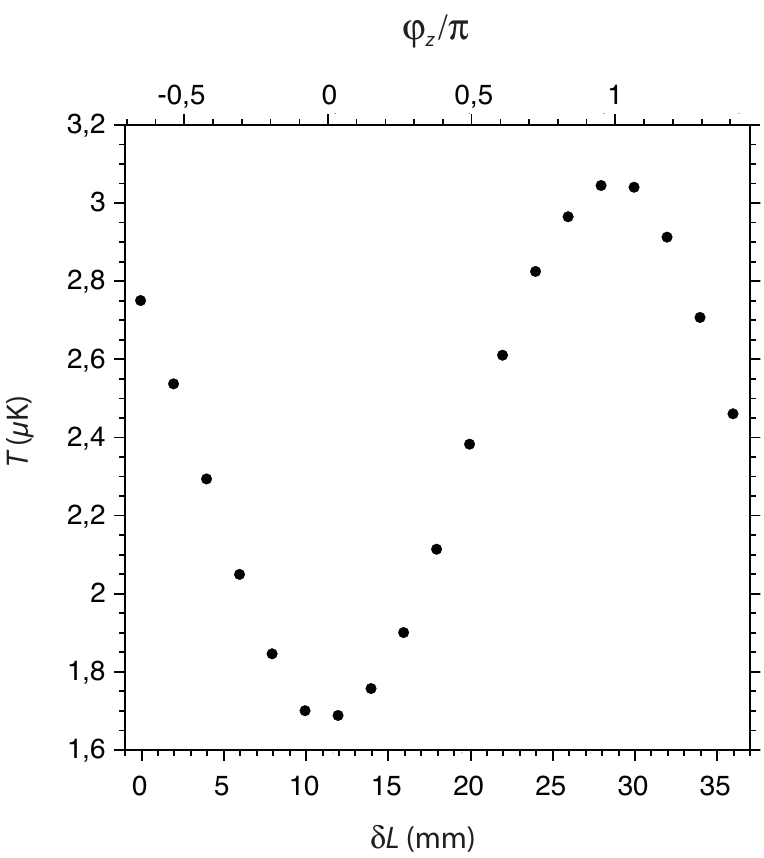}}
\caption{Measured temperature, $T$, dependence as a function of the beam path extension $\delta L$, which determines the conversion between the relative spatial phase to beam path extension \cite{EllmannEPJD}.} 
\label{fig:Temp}
\end{figure}

\begin{figure}
\centerline{\includegraphics[width=0.8\columnwidth]{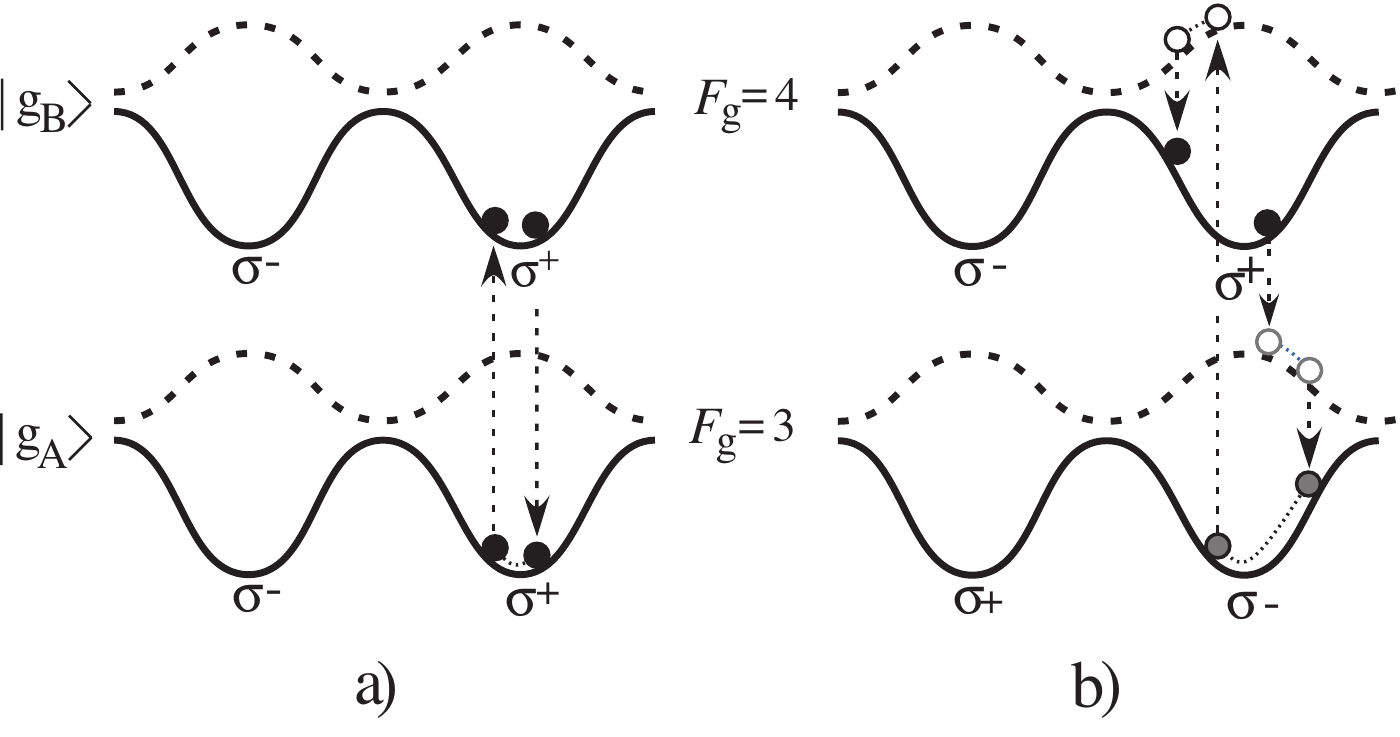}}
\caption{{a) If two $\sigma^+$ points in lattice A (ground state $|\mathrm{g_A}\rangle$) and lattice B (ground state $|\mathrm{g_B}\rangle$) overlap, atoms can be transferred between the lattices with minimised light scattering. b) In the opposite case, when $\sigma^+$  overlaps with $\sigma^-$, heating effects are enhanced.}
\label{semimodel}}
\end{figure}

The temperature dependence is shown in figure \ref{fig:Temp} for a relative spatial phase shift in the vertical $z$-direction. A 3.3 cm extension or shortening of the beam branches corresponds to a relative spatial phase of $2\pi$. This is repeated for the other directions in order to find the origin of all the relative spatial phases. An simplified explanation to the temperature variations \cite{EllmannPRL} is an increase in the number of scattered photons when a $\sigma^+$ point in lattice A overlaps with a $\sigma^-$ point in lattice B. Figure \ref{semimodel}a, shows a semi-classical model where a $\sigma^+$ point in lattice A overlaps with a $\sigma^+$ point in lattice B. If an atom is optically pumped from lattice B to lattice A, it will end up in the lowest light shift potential (due to conservation of angular momentum), and then return to lattice B. In this process, an atom only scatters a few photons. The opposite situation, where a $\sigma^+$ point overlaps with a $\sigma^-$ point is shown in figure \ref{semimodel}b. Here, an atom in lattice B that is pumped to lattice A will end up in the least light shifted anti-trapping potential. The atom will slide down the potential, gaining kinetic energy and, as it is close to a $\sigma^-$ point it will scatter photons, which pumps the atom to the lowest potential. It is then pumped back  to lattice B in a similar process. Thus, in this case the number of scattered photons is considerable higher, which result in heating \cite{EllmannPRL,EllmannEPJD}. In a more realistic picture, there is a number of Zeeman sub-levels, all corresponding to different light shift potentials. For a $\sigma^+$/$\sigma^-$ site overlap, photon scattering increases since atoms are optically pumped between a range of different $M$-states.
\subsection{Measurement of drift velocity by TOF}
\begin{figure}
\centerline{\includegraphics[width=0.7\columnwidth]{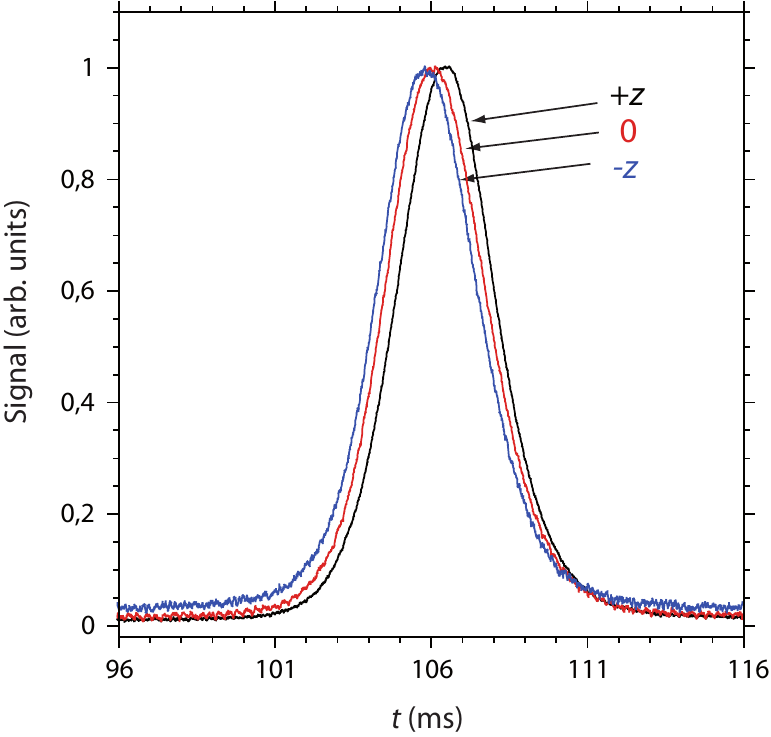}}
\caption{(Colour online) Normalised TOF-signals with arrival time $t$ as the horizontal axis, for relative spatial phases that generates maximum induced drifts in the upward ($+z$) and downward ($-z$)-direction. The middle one ($0$) shows a TOF signal for no induced drifts when the optical lattices are in phase. All three TOF signals are normalised to make comparison easier and the signals are averages of four drops.} 
\label{fig:max_zero_min}
\end{figure}

In figure \ref{fig:max_zero_min}, three TOF signals are shown, all for $\tau=350$ ms. The middle one, indicated by ($0$), is when the two OLs are in phase and no induced drift is present. Those indicated by ($+z$) and ($-z$) are when $\varphi_z$ is set to generate a maximum drift, either upwards or downwards. 
A more direct technique, and a complement to the TOF technique, is imaging the shadow of the atoms, transiently illuminated by a weak resonant probe beam, on a CCD detector, see section \ref{sec:imaging}. The images provides the centre-of-mass position of the atomic cloud in the $xz$ plane, determined by Gaussian fits to the images after an arbitrary lattice time $\tau$.

In order to experimentally investigate the behaviour of the BM, we change the parameters $\Delta_\mathrm{A}$ and $\Delta_\mathrm{B}$, covering the frequency span between the excited $F_\mathrm{e}=3$ to the $F_\mathrm{e}=4$ states, and the $F_\mathrm{e}=4$ and the $F_\mathrm{e}=5$ states respectively. $I_\mathrm{A}$ and $I_\mathrm{B}$ are set independently up to about 20 mW/cm$^2$. To characterise the drift velocity in the vertical $z$-direction ($v_{\mathrm{d}_z}$), we tune the parameters ($\Delta_\mathrm{A}$, $\Delta_\mathrm{B}$, $I_\mathrm{A}$, $I_\mathrm{B}$) in a number of steps, all for a fixed lattice time $\tau$ of 350 ms. For each set of parameters, the relative spatial phase in the vertical $z$-direction ($\varphi_z$) was incrementally changed in about 50 steps, covering slightly more than $2\pi$, while $\varphi_x$ and $\varphi_y$ were kept at zero, meaning that only vertically induced drift was present, either upwards or downwards. In total, we measured about 14000 velocity distributions, due to the large parameter space. For each change of $\varphi_z$, five TOF-signals were averaged to increase the signal-to-noise ratio and Gaussian fits were performed to read out the peak arrival time of the TOF signals.

\begin{figure}
\centerline{\includegraphics[width=0.7\columnwidth]{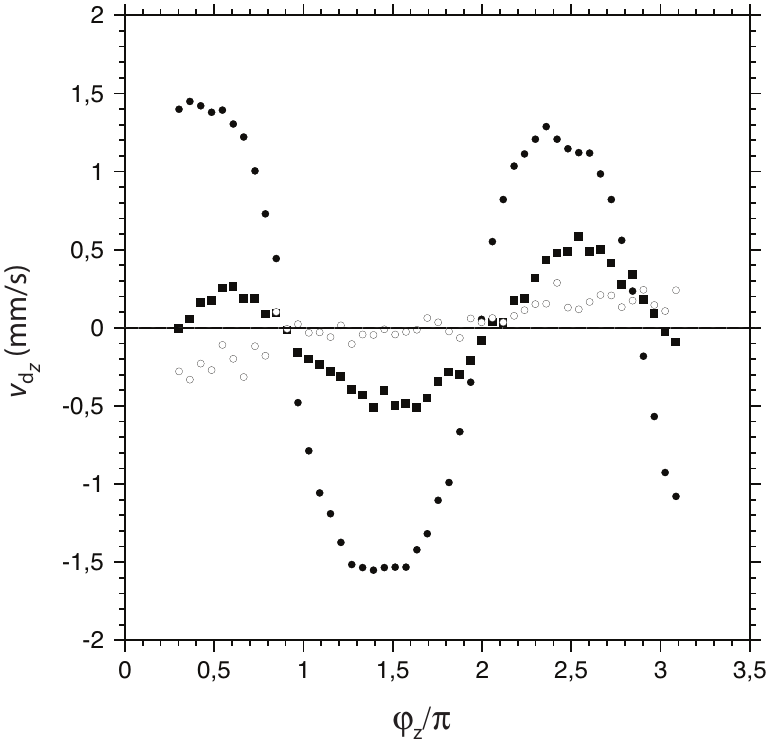}}
\caption{Induced drift velocities $v_{\mathrm{d}_z}$ as a function of relative spatial phase $\varphi_z$ for three different sets of parameters. Filled circles: $\Delta_\mathrm{B}=40\Gamma$, filled squares: $\Delta_\mathrm{B}=25\Gamma$ and circles: $\Delta_\mathrm{B}=10\Gamma$, for a lattice time $\tau$ of 350 ms, $\Delta_\mathrm{B}=14\Gamma$, $I_\mathrm{A}=0.8$ mW/cm$^2$ and $I_\mathrm{B}=1.2$ mW/cm$^2$ for all cases.}
\label{fig:BMcurves}
\end{figure}

\begin{figure}
\centerline{\includegraphics[width=0.7\columnwidth]{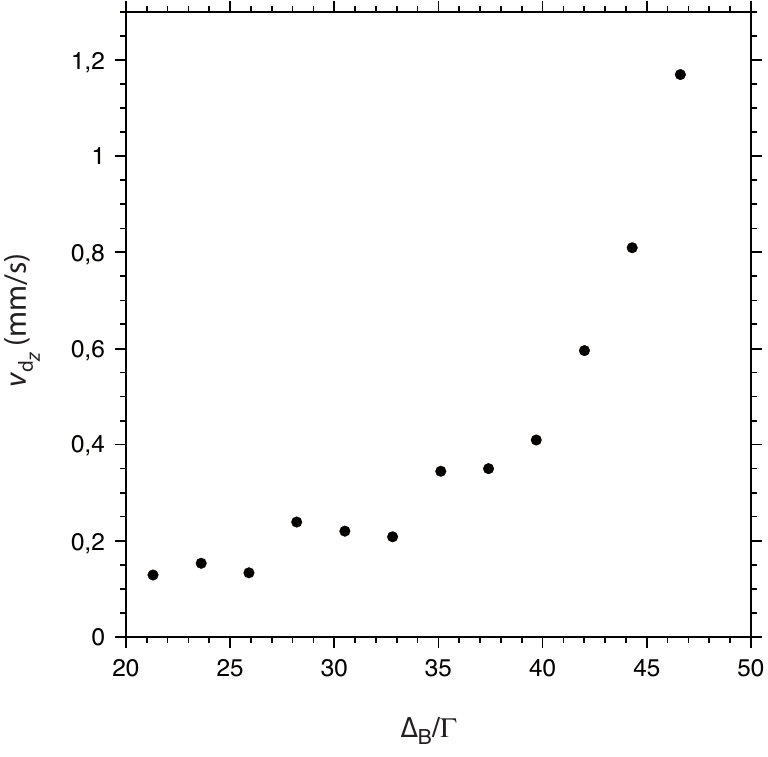}}
\caption{Maximum vertical drift velocity $v_{\mathrm{d}_z}$ as a function of $\Delta_\mathrm{B}$, where $\Delta_\mathrm{A}=33\Gamma$, $I_\mathrm{A}=4.17$ mW/cm$^2$, $I_\mathrm{B}=2.71$ mW/cm$^2$ and the lattice time $\tau=450$ ms. The $F_\mathrm{g}=4\rightarrow F_\mathrm{e}=4$ resonance occurs at $\Delta_\mathrm{B}=48\Gamma$.}
\label{fig:varyD45}
\end{figure}

\begin{figure}
\centerline{\includegraphics[width=0.7\columnwidth]{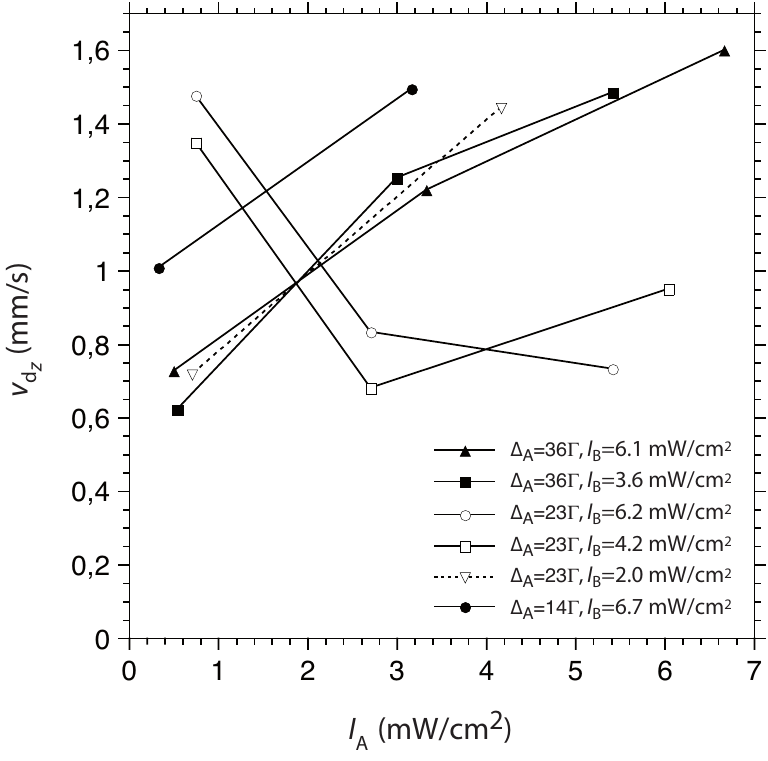}}
\caption{Maximum drift velocity $v_{\mathrm{d}_z}$ in the ($+z$)-direction for different sets of parameters, plotted as a function of $I_\mathrm{A}$. Meanwhile, $\Delta_\mathrm{B}=40\Gamma$ is kept fixed during a lattice time $\tau=350$ ms.} 
\label{fig:strange}
\end{figure}

With a constant drift velocity \cite{PederPRL}, the vertical drift velocity $v_{\mathrm{d}_z}$ is determined from the peak arrival time $t$ of the atoms at the TOF probe by 
\begin{equation}
v_{\mathrm{d}_z}=\frac{gt^2-2l}{2(t+\tau)},
\end{equation}
\label{eq:conversion}
where $l$ is the vertical distance that the atoms fall down to the probe and $g$ is the gravitational acceleration. In figure \ref{fig:BMcurves}, three different curves are plotted, showing $v_{\mathrm{d}_z}$ as a function of $\varphi_z$ for three different values of $\Delta_\mathrm{B}$, while $\Delta_\mathrm{A}$, $I_\mathrm{A}$ and $I_\mathrm{B}$ are kept fixed. Here, a clear variation of $v_{\mathrm{d}_z}$ is evidenced for different values of $\Delta_\mathrm{B}$. In figure \ref{fig:varyD45}, $\Delta_\mathrm{B}$ is plotted as a function of $v_{\mathrm{d}_z}$ for a relative spatial phase that generates a maximum $v_{\mathrm{d}_z}$ in the upward ($+z$)-direction, with $\Delta_\mathrm{A}$, $I_\mathrm{A}$ and $I_\mathrm{B}$ fixed for a lattice time $\tau$ of 450 ms. The maximum drift velocity clearly increases when $\Delta_\mathrm{B}$ increases and approaches the $F_\mathrm{e}=4$ state. Continuing the increase of $\Delta_\mathrm{B}$ close to the $F_\mathrm{e}=4$ state results in an abrupt decrease of the TOF-signal due to resonant trap losses. However, changing $\Delta_\mathrm{A}$, $I_\mathrm{A}$ and $I_\mathrm{B}$, while keeping $\Delta_\mathrm{B}$ fixed, affect $\varphi_z$ less  significantly and in a more ambiguous way compared with changing $\Delta_\mathrm{B}$, as can be seen in figure \ref{fig:strange}. With the large parameter space and with the high degree of coupling between the different parameters, it is difficult to extract any unambiguous trends from the data.
The behaviour of this type of BM is complex and since it is working in the dissipative regime, a strong coupling between friction, heating and dissipation is present. Changing any parameter, simultaneously changes the friction, heating and dissipation in the system as well as the transfer rates.
Basically, since the rectification mechanism emanates from the asymmetry in the transfer rates between the two OLs, it seems that the asymmetry conditions for larger induced drifts generally increase when approaching the $F_\mathrm{e}=4$ resonance. When detuning closer to the $F_\mathrm{e}=4$ transition for lattice B, the possibility for optical pumping of atoms to the $F_\mathrm{g}=3$ ground state due to absorption of non-resonant photons increases, which in turn seem to enlarge the BM effect since atoms are spending longer time in the short lived optical lattice A. Note that the optimal asymmetry in the transfer rates, maximising the induced drift, does not correspond to the maximal asymmetry \cite{Laurent}. Typically, the ratio between the transfer rates $\gamma_\mathrm{A\rightarrow B}:\gamma_\mathrm{B\rightarrow A}$ are in the order of 9:1, which is due to the fact that one is closed and one is an open transition \cite{EllmannEPJD}.
\subsection{Evidence for horizontal and vertical drifts}
\label{sec:imaging}
\begin{figure}
\centerline{\includegraphics[width=0.8\columnwidth]{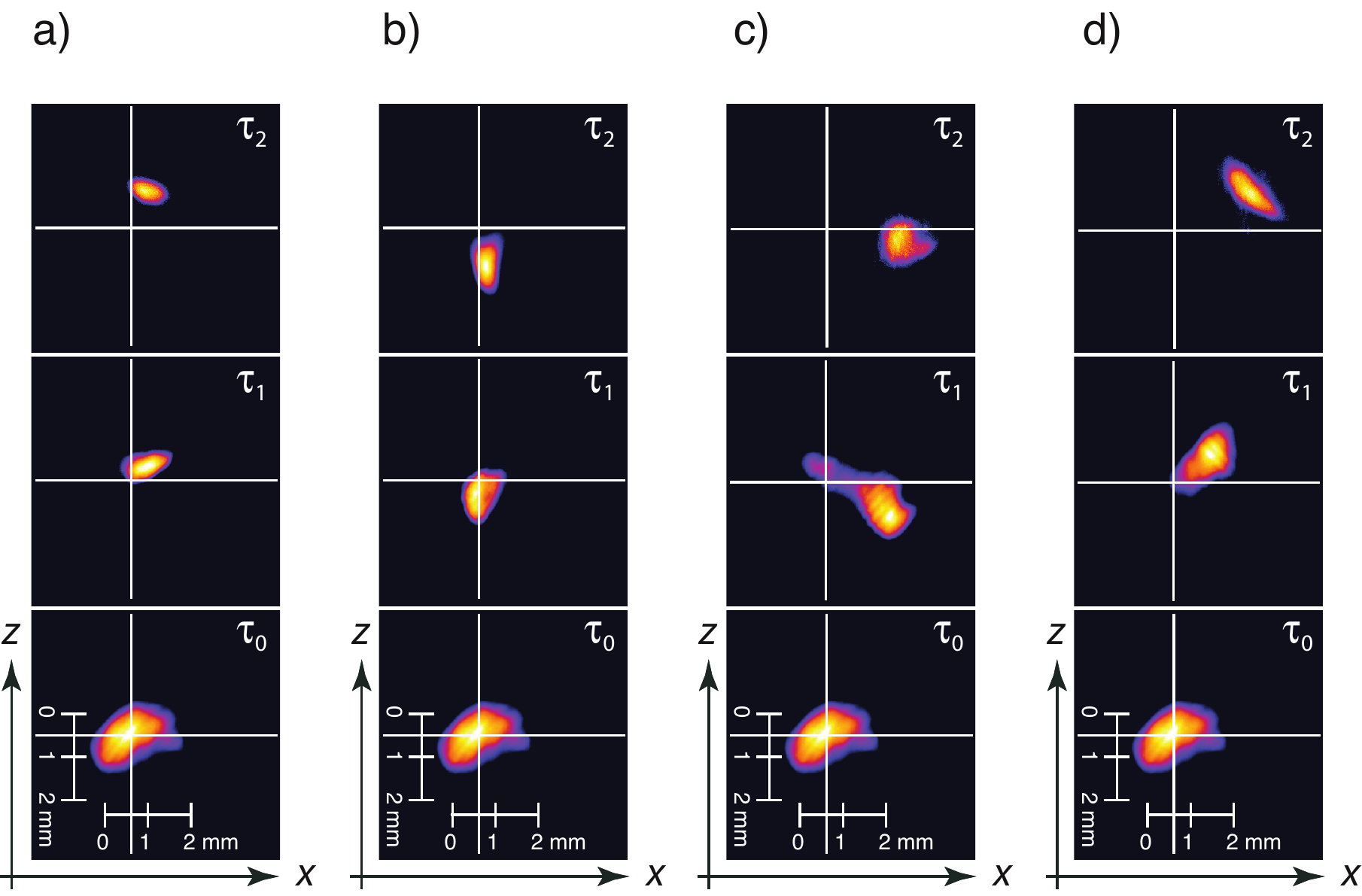}}
\caption{(Colour online) False colour images of the atomic cloud at two different interaction times $\tau_\mathrm{1}$ and $\tau_\mathrm{2}$, $\tau_\mathrm{0}$ being with no interaction time ($\tau=0$ s). The final image is acquired after an interaction time of $\tau_\mathrm{2}\sim$0.9 s ($\tau_\mathrm{1}\sim$0.5 s). \textbf{a)} A phase shift only along $z$, showing an upward drift. \textbf{b)} Different phase shift in $z$. A downward drift is evident, $\varphi_x$ and $\varphi_y$ still being zero. \textbf{c)} $\varphi_z$ and $\varphi_y$ are zero. Only a horizontal drift along $x$ is evident. \textbf{d)} Both $\varphi_x$ and $\varphi_z$ are set for a maximum drift in both the $z$ and $x$-directions, resulting in a diagonal drift.} 
\label{fig:camera}
\end{figure}

\begin{figure}
\centerline{\includegraphics[width=0.7\columnwidth]{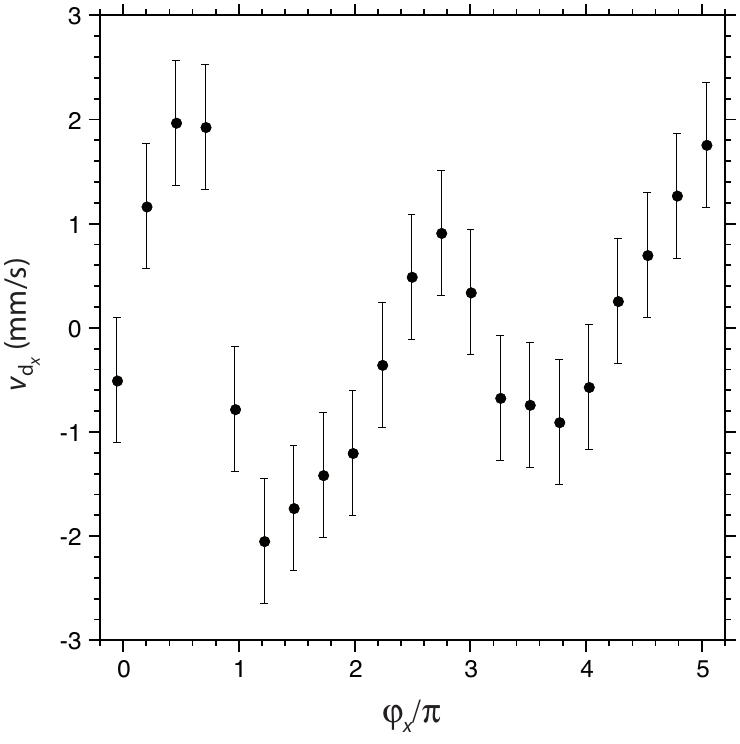}}
\caption{Experimental results showing the drift velocity $v_{\mathrm{d}_x}$ as a function of relative spatial phase along $x$. Here, $\Delta_\mathrm{B}=40\Gamma$, $\Delta_\mathrm{A}=36\Gamma$, $I_\mathrm{B}=6.13$ mW/cm$^2$ and $I_\mathrm{A}=6.67$ mW/cm$^2$ for a lattice time $\tau$ of 350 ms. The error bars come from the uncertainty in the Gaussian fits used to determine the centre-of-mass $x$-position to the images.} 
\label{fig:vx_and_vz}
\end{figure}

Our lattice structures are periodic in three dimensions. We can adjust the relative spatial phases $\varphi_z$, $\varphi_y$ and $\varphi_x$ at will. Thus, the Brownian motor works also horizontally, and indeed in an arbitrary direction. To confirm this we measure the position of the atomic cloud in the $xz$-plane as a function of the lattice time $\tau$. This is done by imaging the shadow of the atomic cloud, transiently illuminated by a weak probe beam, on a CCD-detector. Figure \ref{fig:camera} shows false colour images of the atomic cloud for different successive lattice times. An induced drift is evident both along $z$ and $x$, and as well as in the diagonal $xz$-direction for an appropriate choice of $\varphi_x$ and $\varphi_z$ ($\varphi_y=0$) which generates the largest drift velocities while keeping $\Delta_\mathrm{B}$, $\Delta_\mathrm{A}$, $I_\mathrm{B}$ and $I_\mathrm{A}$ fixed. 
To confirm the induced drift dependence along $x$ as a function of $\varphi_x$, we determine for each change in $\varphi_x$ the centre-of-mass $x$-position by a Gaussian fit to the images, while $\varphi_z$ and $\varphi_y$ where kept fixed at zero. For comparison reason with TOF results in the $z$-direction, the same lattice time $\tau=350$ ms was used for one typical set of parameters, see figure \ref{fig:vx_and_vz}.
\section{Numerical simulations}
\label{sec:Classical simulations}
\begin{figure}
  \centerline{\includegraphics[width=0.8\columnwidth]{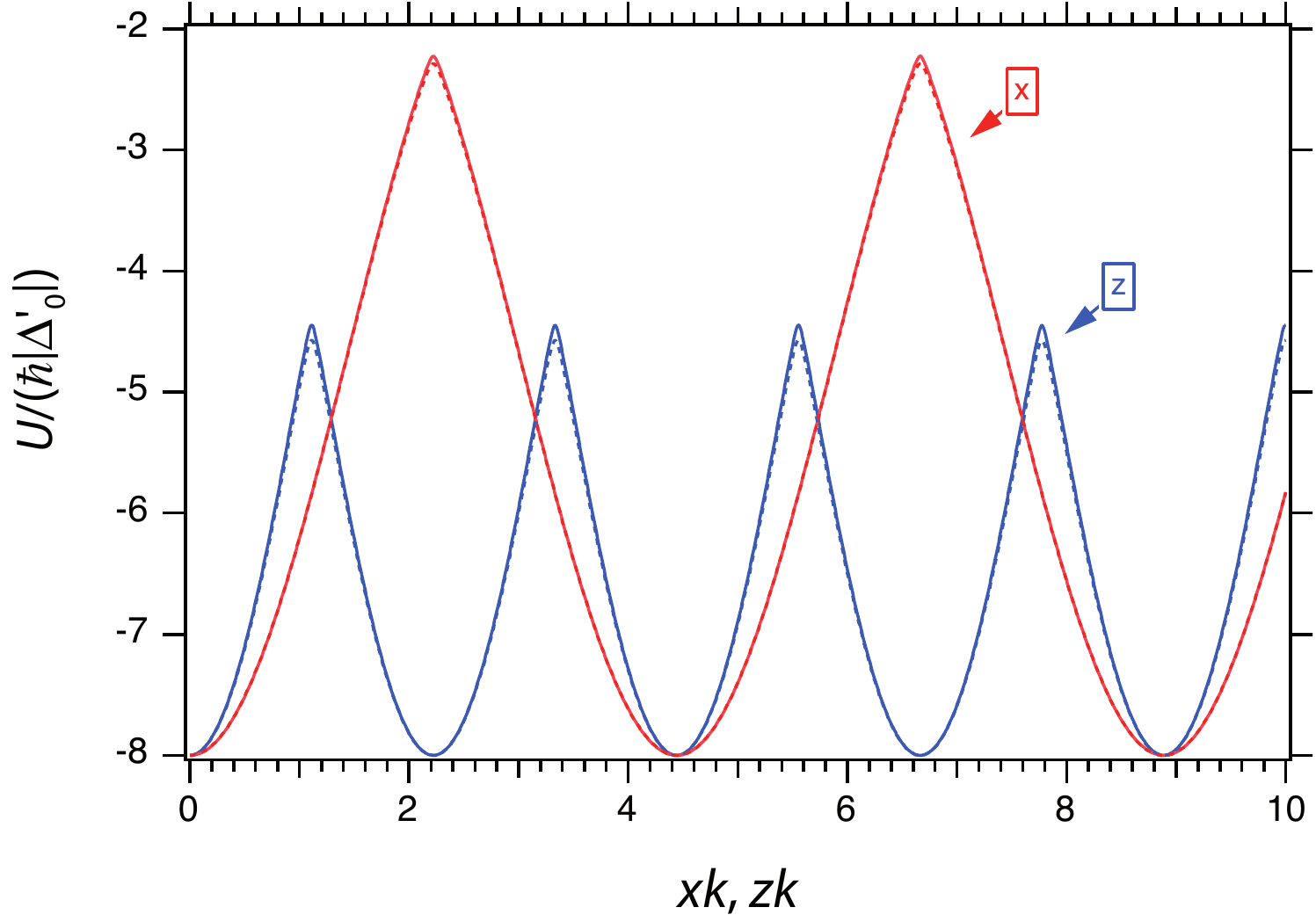}}
  \caption{(Colour online). One dimensional cuts $U(0,0,z)$ (blue) and $U(x,0,0)$ (red) of the potential surfaces used in the simulation, expressed as a function of the light shift $|\Delta_0'|$ \cite{grynberg01}, where $k$ is the absolute value of the wavevector of the laser field.  Solid line: lowest adiabatic potential of the $F_\mathrm{g} = 4 \rightarrow F_\mathrm{e} = 5$ transition; dashed line: lowest adiabatic potential of the $F_\mathrm{g} = 3\rightarrow F_\mathrm{e} = 4$ transition.}
  \label{fig:potentials}
\end{figure}

In order to understand the qualitative behaviour of our BM, we have
performed simulations, using a simple model displaying the basic
characteristics of our system. We consider a classical Brownian
particle which can be in one of two internal states, indexed by
\textit{j}, and interacting with state-dependent external potentials
$U_j$.  In analogy to the experimental setup, section
\ref{sec:Experiment}, we chose $U_\mathrm{B}$ to be the light shift
potential resulting from laser field B (see figure~\ref{fig:levels}),
corresponding to the lowest adiabatic optical potential
~\cite{grynberg01} for the $F_\mathrm{g} = 4 \rightarrow F_\mathrm{e}
=5$ transition of the 4-beam lin$\perp$lin configuration.
Consequently, $U_\mathrm{A}$ is chosen as the lowest adiabatic optical
potential for the $F_\mathrm{g} = 3 \rightarrow F_\mathrm{e} = 4$
transition.  These potentials are illustrated in
figure~\ref{fig:potentials},
where the difference in periodicity and height between the $x$ and $z$
directions should be noted (cuts along $x$ or $y$ are identical),
while $U_\mathrm{A}$ and $U_\mathrm{B}$ are almost the same. We
consider unequal transfer rates $\gamma_{\mathrm{A}\rightarrow
  \mathrm{B}}\neq \gamma_{\mathrm{B} \rightarrow \mathrm{A}}$ between
the two potentials. The Brownian motion is governed by the scattering
of photons from the optical lattice lasers, represented by the
diffusion constants $D_{\mathrm{v},j}(\textbf{x})$, while laser
cooling provides friction. The Fokker-Planck equation for the Wigner
distribution $W_j(\textbf{x},\textbf{v},t)$ of this Brownian particle,
characterised by its position $\textbf{x}$, velocity $\textbf{v}$ and
internal state \textit{j} as A or B, reads~\cite{Laurent,Risken}
\begin{multline}
  \left[ \partial_t + v \partial_{\mathbf{x}} \right]W_j - \partial_{\mathbf{v}} \left[ \mathbf{v} + \mbox{\boldmath $\nabla$}\mathbf{U}_j(\mathbf{x}) + D_{\mathrm{v},j}(\mathbf{x}) \partial_{\mathbf{v}} \right] W_j  \\ 
  = \gamma_{j' \rightarrow j}(\mathbf{x}) W_{j'} - \gamma_{j \rightarrow j'}(\mathbf{x}) W_j \label{eq:FPE} 
\end{multline}
for $j' \neq j$, with time expressed in units of the inverse of the
friction coefficient $\alpha$ and position in units of the inverse
angular wave vector $\vec{k}$ of the laser light, such that all
variables are dimensionless.  In addition, we scale the potentials by
a factor of $(45/88)\mathcal{A}$,\footnote{A coefficient of 44/45
  appears in the formula for the potential from the Clebsch-Gordan
  coefficients for a $F_{\mathrm{g}} = 4 \rightarrow F_{\mathrm{e}} =
  5$ transition.} such that the adiabatic potential along $z$,
$U(0,0,z)$ shown in figure~\ref{fig:potentials}, has a depth of
$\approx 2\mathcal{A}$, allowing a direct comparison with the sine
potential amplitude $\mathcal{A}$ used in ref.~\cite{Laurent}. We
assume that the mean kinetic energy of the system is smaller than the
depth of the trapping potentials and that the typical frequency of the
oscillations in the potential wells is larger than the rate of energy
damping, so that the dynamics falls into the low-damping regime.
Motion is restricted to 2D ($y=0$) and we observe the average velocity
$v_\mathrm{d}$ in the $z$ or $x$ directions as a function of the
relative phase shift $\varphi$ in either of these directions.  Unless
noted otherwise, we use arbitrary position-independent values of
$D_\mathrm{v}$, and $\gamma$, with $\gamma_{\mathrm{A} \rightarrow
  \mathrm{B}} = 3\gamma_{\mathrm{B} \rightarrow \mathrm{A}} = 7.5$
(\textit{i.e.}, an atom spends 75\% of its time in the long-lived
lattice B).

\begin{figure}
  \centerline{\includegraphics[width=0.8\columnwidth]{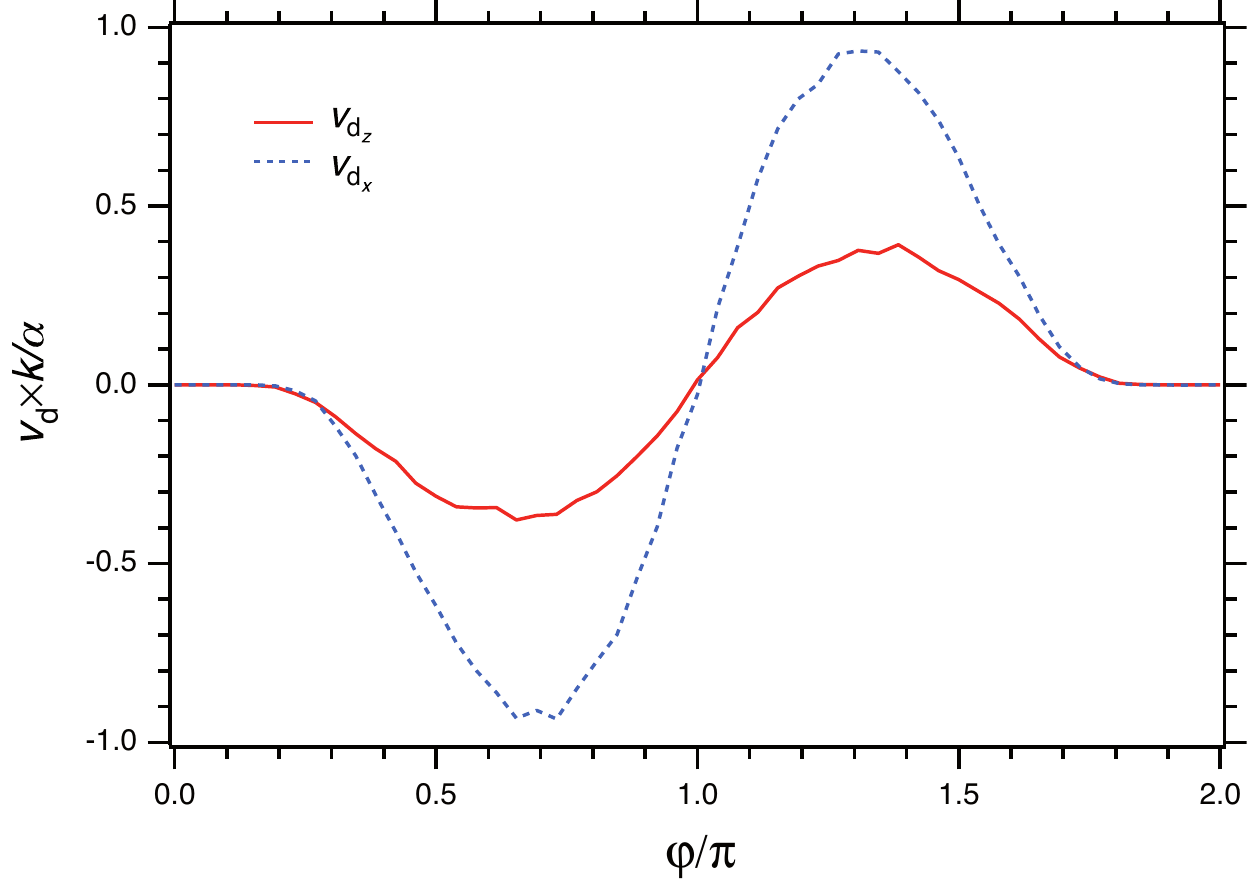}}
  \caption{Results from numerical simulations of the Brownian motor mechanism for the vertical $z$ (full line) and horizontal $x$ (dashed line) directions. The drift velocity $v_\mathrm{d}$ is plotted as a function of the relative spatial phase $\varphi=\varphi_\mathrm{A}-\varphi_\mathrm{B}$, varied independently along $z$ or $x$.} 
\label{fig:vx_and_vy-sim}
\end{figure}

\begin{figure}
  \centerline{\includegraphics[width=0.8\columnwidth]{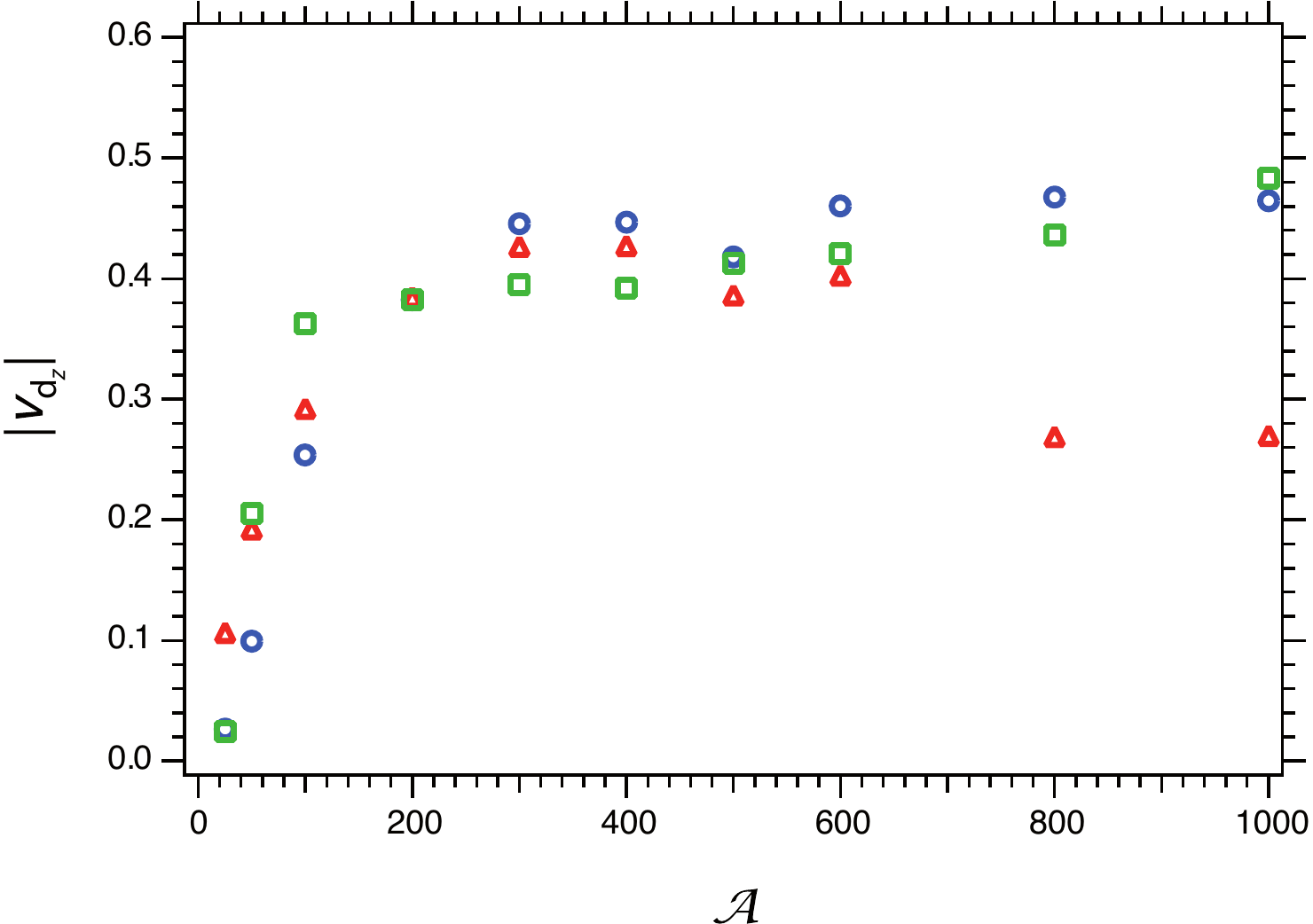}}
  \caption{Dependence of the magnitude of drift velocity on the depth of the potentials.  Circles: identical depth $\mathcal{A}_{\mathrm{A}} = \mathcal{A}_{\mathrm{B}}$; squares: varying depth of the long-lived lattice $\mathcal{A}_{\mathrm{B}}$ for $\mathcal{A}_{\mathrm{A}} = 200$; triangles: varying depth of the short-lived lattice $\mathcal{A}_{\mathrm{A}}$ for $\mathcal{A}_{\mathrm{B}} = 200$.  $\varphi_z = 2\pi/3$ and $D_\mathrm{v} = 75$ in all cases.} 
\label{fig:simulA}
\end{figure}

\begin{figure}
  \centerline{\includegraphics[width=0.8\columnwidth]{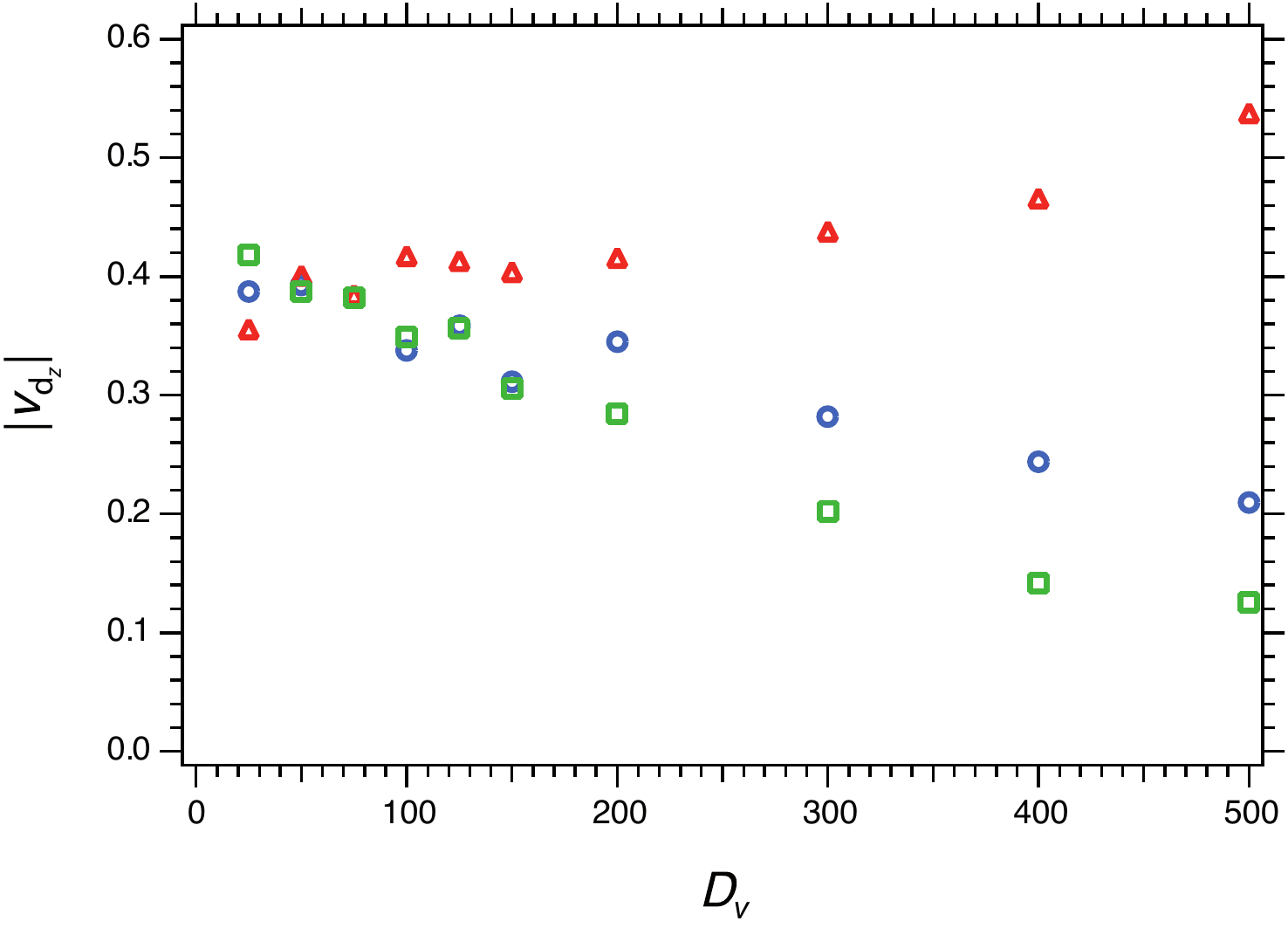}}
  \caption{Dependence of the magnitude of drift velocity on the diffusion constant. Circles: identical diffusion $D_{\mathrm{v,A}} = D_{\mathrm{v,B}}$; squares: varying diffusion in the long-lived lattice $D_{\mathrm{v,B}}$ for $D_{\mathrm{v,A}} = 75$; triangles: varying diffusion in the short-lived lattice $D_{\mathrm{v,A}}$ for $D_{\mathrm{v,B}} = 200$.  $\varphi_z = 2\pi/3$ and $\mathcal{A} = 200$ in all cases.} 
\label{fig:simulDv}
\end{figure}

Figure~\ref{fig:vx_and_vy-sim} shows the dependence of the Brownian motor mechanism on the relative phase $\varphi$ between the two potentials, 
with $\mathcal{A} = 200$ and $D_{\mathrm{v}} = 75$ in both states. The overall shape of the curve is similar to what was obtained for a sine potential in ref.~\cite{Laurent}, with a greater drift velocity for a phase shift along $x$, due to the deeper potential along that direction. Indeed, figure~\ref{fig:simulA} shows that the drift velocity is proportional to the potential depth, up to a certain value where the BM effect saturates. 
Similarly, changing the diffusion constant $D_{\mathrm{v}}$ affects the BM differently depending on the internal state of the atom, as seen in figure~\ref{fig:simulDv}. 
Increasing the diffusion in both lattices, or only in the long-lived
one, results in a decrease of the drift velocity, due to the added
noise.  Conversely, a greater diffusion in the short-lived lattice
slightly increases the drift velocity, because the atom can take
advantage of the increased Brownian motion without too much adverse
effect from isotrope diffusion. 

Compared to the simulations in figure \ref{fig:vx_and_vy-sim}, the
experimental results clearly deviate around $\varphi=0$ and $2\pi$
both in $x$ and $z$.  This was also noticed in simulations using
diabatic potential curves~\cite{DionEPJST}, which prompted us to use
adiabatic curves in the present study.  It is now clear that the
discrepancy is not due to the shape of the potential, which calls for
simulations where a more realistic model is used to approach the
physical system at hand.  Moreover, we note that the relation between
the beam path extension and the relative spatial phase, shown in
figure~\ref{fig:Temp}, indicates that Brownian motor effect has the
same periodicity as the diabatic potential curve~\cite{EllmannEPJD}.
\section{Discussion}
\label{sec:Discussion}
There are two ambient forces that may affect the experimental results. One is due to earth magnetic field, from which a slight Zeeman shift could be introduced. This is cancelled by a B-field compensation in the experimental setup. The second force is due to gravity. It will introduce a slight tilt of the lattice potentials, which may results in an increased tunnelling probability in one direction. This effect is small, and for potential depths used in the experiment of 100--300$E_\mathrm{r}$. The potential energy $U_\mathrm{p}=mgh$ contribution due to gravity becomes $\sim 10^{-3} E_\mathrm{r}$, where $m$ is the atomic mass, $g$ is the gravitational constant and $h=\lambda/\sqrt{2}$ is the distance between two $\sigma^+$ $\sigma^-$-sites in the vertical z-direction ($\lambda=852.3$ nm). This clearly indicates that this contribution is negligible compared to the potential depth.\\

While two periodic potentials are used in the model, in reality the atoms have magnetic substates, each leading to a different light shift potential. A further difficulty with our BM is that, since it works in a dissipative regime, the parameters, potential depths, magnitude of the diffusion, and the transition rates between the two lattices all depend on the irradiances and the detunings of the OL lasers. Changing any parameter results in a different friction, diffusion and as well as dissipation. Therefore, one of the main features of this paper is to investigate how this type of Brownian noise rectifier varies under different conditions. The complexity of the system makes it difficult to directly relate the experimental findings to the simple classical model used in the simulation. Quantum mechanical simulations in terms of parameters directly controllable in experiments, using the full level structure, are under development.

Despite these difficulties, carefully investigations of the parameter space, $\Delta_\mathrm{A}$, $\Delta_\mathrm{B}$,  $I_\mathrm{A}$ and $I_\mathrm{B}$, allow some general conclusions to be drawn. The most striking feature is the drastic increase of the effectiveness of the BM when $\Delta_\mathrm{B}$ is increased. As $\Delta_\mathrm{B}$ is increased, the frequency of the light will get closer to the $F_\mathrm{g}=4\rightarrow F_\mathrm{e}=4$ transition, and hence the pumping rate $\gamma_{\mathrm{B\rightarrow A}}$ increases. Classical simulations \cite{Laurent} confirm that there is an optimal ratio between pumping rates, as long as the inequality remains large. Changing $\Delta_{\mathrm{B}}$ will also change the pumping rate between the sublevels {\em within} the lattice B manifold. Therefore, the potential depth, friction and diffusion within lattice B will also be modified, both in magnitude and position dependence, in a non-straightforward way \cite{EllmannEPJD2001}.

However, in the limit $\gamma_{\mathrm{B \rightarrow A}}/\gamma_{\mathrm{A \rightarrow B}}\rightarrow 0$ or $\infty$ the system is effectively reduced to a single optical lattice, and we would expect any BM effect to vanish. Therefore it is clear that the optimal BM effect must be achieved for some finite ratio $\gamma_{\mathrm{B\rightarrow A}}/\gamma_{\mathrm{A\rightarrow B}}$, as has indeed been confirmed by classical simulations \cite{Laurent}. This is clearly consistent with the data in Fig.~\ref{fig:varyD45}. Due to experimental difficulties when laser B is tuned too close to level A we have not been able to extend our data to larger $\Delta_\mathrm{B}$ in order to investigate if the drift velocity falls off again. Classical simulations indicate that this will hapen for $\gamma_{\mathrm{B \rightarrow A}}/\gamma_{\mathrm{A \rightarrow B}}$ larger than 0.1-0.2. Although we do not know the exact relation between $\gamma_{\mathrm{B \rightarrow A}}/\gamma_{\mathrm{A \rightarrow B}}$ and $\Delta_\mathrm{B}$ (which will also depend on other parameters) measurements of the  relative populations of the two lattices indicate that this value has not been reached \cite{EllmannEPJD}.

For quantitative analyses, the model using only two potentials, while ignoring the internal level structure of the two lattices, is not sufficient. This is evident since features like the ones at $\varphi=0$ and at $\varphi=2\pi$ for the directional variation of the shape of the velocity curve in figure \ref{fig:vx_and_vy-sim} do not show up in the experimental graphs (see figure \ref{fig:BMcurves}). It is also expected that the position dependence of pumping, friction and diffusion coefficients will greatly affect the shape of the curve.
\section{Conclusion}
\label{sec:Conclusion}
In summary, we have demonstrated a Brownian motor working in three dimensions with a controllable speed. Induced drift velocities in the order of one recoil velocity\footnote{One recoil velocity $v_\mathrm{r}$ is about 3 mm/s for a Cs atom scattering a lattice photon.} have been achieved. We have showed that some of the qualitative features of the experiment can be reproduced theoretically using a classical model similar to that in \cite{Laurent}, which has been extended to two dimensions.

Up to now, a wide variety of BMs have been theoretically investigated and also demonstrated for various systems \cite{Reimann,Astumian}. The main features of our BM can be qualitatively described using a purely classical model shown in section \ref{sec:Classical simulations}. Nevertheless, the coupling between the potentials is driven by quantum jumps as resulting from spontaneous emission, which essentially is a quantum mechanical feature. Together with quantised motion, this may open the way for creation of a quantum Brownian motor \cite{Hanggi_quantum}. Due to the generality of our scheme, applicability to chemical and/or biological systems may also be possible.
\section{Acknowledgements}
We thank M. Nyl\'{e}n and L. Sanchez-Palencia for helpful discussions. This work was supported by Knut och Alice Wallenbergs stiftelse, Vetenskapsr{\aa}det, Carl Tryggers stiftelse, Kempestiftelserna. Part of this research was conducted using the resources of the High Performance Computing Centre North (HPC2N).
\bibliographystyle{epj}
%\bibliography{bibfiles/shortnames,bibfiles/references}

\end{document}